\documentclass[rmp,letterpaper,twocolumn,noshowpacs,preprintnumbers,amsmath,amssymb,showkeys,longbibliography]{revtex4-1}

\usepackage{epsfig}
\usepackage{graphicx}
\usepackage{dcolumn}
\usepackage{bm}
\usepackage{bbm}
\usepackage{color}

\begin{document}


\title{Semi-Analytical Pricing for General Default Intensity Models}

\author{Ryan Parker$^{1}$} \author{Mark Stedman$^{2}$}  \author{Luca Capriotti$^{3}$}

\affiliation{%
$^{1}$ The University of Cambridge,  Cambridge CB3 0HE, United Kingdom \looseness=-10\\
$^{2}$ Jain Global LLC, 9 West 57th Street, New York, New York 10019, United States of America \looseness=-10\\ 
$^{3}$ Columbia University, New York, New York 10027, United States of America \looseness=-10 
}%

\date{\today}

\begin{abstract}
Using the path-integral formalism, we develop an accurate and easy-to-compute semi-analytical approximation for a general class of {default intensity} models. 
We illustrate the accuracy of the method by presenting results for the Black-Karasinski model for which the proposed approximation provides remarkably accurate results, 
even in regimes of high volatility and multi-year time horizons. The accuracy and the computational efficiency of the proposed approximation makes it a viable alternative 
to fully numerical schemes for a variety of applications in econometrics and derivatives pricing, including the computation of XVA for credit products.  
As a practical example,  we consider the pricing of a quanto Credit Default Swap (CDS) under stochastic intensity of default and an FX devaluation model.
\end{abstract}


\maketitle

Default intensity models are of paramount importance in financial modeling, providing the foundation of many approaches 
used for pricing credit derivatives.

In this paper, we consider a class of realistic models 
in which 
the default intensity assumes the form $h_t = h(X_t)$  with $X_t$ following the non-linear diffusion process  \begin{equation}\label{eq.diffusion}
dX_t =k ( \theta_{t} - X_t)\,dt + \sigma \,dW_t~,
\end{equation}
where $\theta_{t}$ is a time-dependent mean reversion level, $\sigma$ is the  volatility, $X_{0} = x_{0}$, and $W_t$ is a standard Brownian motion.  
The \citep{hull1990pricing} and \cite{bk} (BK) models, where $h(X_t) = X_{t}$ and  $h(X_t) =\exp{X_t}$, respectively, are conspicuous examples of such general
models.

BK is particularly suitable for 
credit modelling
as it ensures that the default intensity is
positive.  Further, the choice of constant $k$ and $\sigma$ parameters is justified in this
context since the credit volatility market is typically liquid at only a few short-dated option
expiries.

Unfortunately, BK is not analytically tractable. For this reason, although widely used in practice, BK implementations rely on computationally intensive  partial differential equation  (PDE) or Monte Carlo (MC) methods for the calculation of 
survival probabilities, which are the essential building blocks for derivatives pricing.

The problem of finding analytic expressions for the BK model has been considered by several authors, starting with \cite{Hagan07, daniluk2016, stehlikova2014effective}, who addressed it using expansions in time to maturity or variance.
In another Risk paper, \cite{Mercurio2019} applied the so-called chaos expansion 
to a mean-reverting stochastic process similar to the BK model. More recently, \cite{horvath2018analytic} proposed a different approach, assuming that deviations of the default intensity from the forward curve are on average small in absolute terms. They found that the required Arrow-Debreu (AD) density can be written as an asymptotic power series.
The approach was extended to the calculation of option prices in \cite{turfus2020perturbation}.


In this paper, we show how the path integral formalism can be used to develop accurate approximations for survival probabilities and AD densities for SDEs of the form  (\ref{eq.diffusion}), even in high-volatility regimes, which are especially relevant for default intensity modeling.


We start with the generalized AD densities,  also known as Green's functions. 
These are defined, in this setting, as 
\begin{equation} \label{eq.ad}
\psi_\lambda(x_T, x_{0}, T) = \mathbb{E}\Big[\delta(X_{T}-x_T)e^{-  \lambda \int_{0}^T du \,h_u }\Big],
\end{equation}
where $\lambda$ is a real number,  and $\delta(\cdot)$ is the standard Dirac's delta function.
This, for $\lambda =0$, gives the transition density, which is central to econometric applications, 
while the price at time $t=0$ of a European option with expiry $T$ and payout of the form $P(h_T)$, 
\begin{equation}
V(0) = \mathbb{E}\Big[e^{-\int_{0}^T du \, h_u} P(h_T) \Big],
\end{equation}
can be obtained by integrating  the product of the payout function and the ($\lambda = 1$) AD density over all the possible values of the default intensity
at time $T$, namely
\begin{equation}\label{eq.cc}
V(0) = \int dx_T\, \psi_1(x_T, x_{0}, T) P(x_T) ,
\end{equation}
where the integration is performed over the range of the function $x_T=h^{-1}(h_T)$ and, with a slight abuse of notation, we denote {$P(x_T) = P ( h(x_T) )$}. In particular, setting $P\equiv 1$
gives the value at time $t=0$ of 
the survival probability up to time $T$, conditional on survival up to time $t=0$,
\begin{equation}
Q(x_{0}, T) = \mathbb{E}\left[1_{\{\tau>T\}}| x_{0}, \tau > 0\right]=\mathbb{E}\left[e^{-\int_0^T du \,h_{u}}\right]~.
\label{eq.survprob}
\end{equation}

%
%

\begin{figure}[t]
\centerline{\includegraphics[width=70mm]{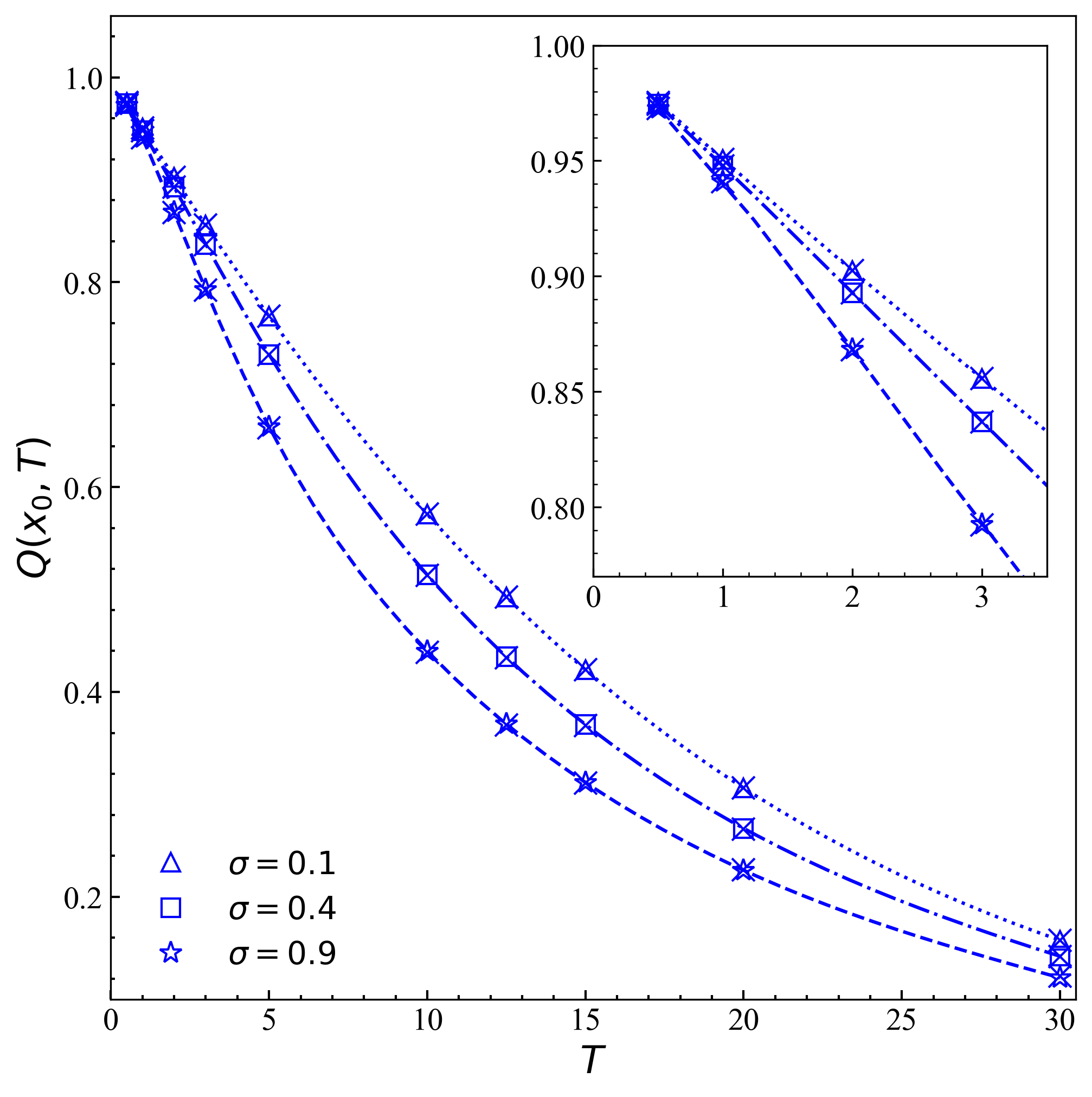}}
\vspace*{8pt}
\caption{Comparison of GTFK and PDE survival probabilities as a function of time to maturity for the Black-Karasinski model, with mean reversion speed $k = 0.1$, initial rate $h_{0} = 0.05$, different values of the volatility, and piecewise constant mean reversion levels of the form explained in the text, with $\theta_{min} = \log(0.05)$ and $\theta_{max} = \log(0.10)$ and $N_{l}=10$.  Crosses indicate the PDE results. The inset is an enlargement for short times to maturity.}
\label{zeroBK}
\end{figure}


A path-integral representation of the AD density can be constructed as follows
\begin{equation} \label{eq.ADprice}
\psi_\lambda(x_T,  x_0, T)  =  e^{-W(x_T, x_0)}  \rho(x_T,  x_0, T)~,
\end{equation}
where 
\begin{equation}\label{eq.p2i}
\rho(x_T,  x_0, T) = \int_{x(0) =x_0}^{x(T) = x_T} \hspace{-0.1cm}{\cal D} [x(t)] \, \,e^{S_{\rho}[x(t)]}~
\end{equation}
is the {\em density matrix}, 
the functional
\begin{equation}\label{eq.action}
S_{\rho}[x(t)] = - \int_{0}^T dt \left[ \frac{1}{ 2\sigma^2 } \dot x^2(t) + V(x(t),t) \right]~
\end{equation}
is the {\em Euclidean action}, 
\begin{equation}\label{eq.driftpot}
V(x, \dot x, t) = \frac{k^{2}(\theta_{t} - x)^2}{2\sigma^2} + \frac{k}{\sigma^2} \dot \theta_{t}  x   + \lambda r(x)~
\end{equation}
is the {\em drift potential}, and we have defined
\begin{eqnarray*}
W(x_T, x_0) = \frac{k}{\sigma^{2}} \left[ \frac{x_{T}^{2}-x_{0}^{2}}{2} - \Big(\theta_{T} x_{T} - \theta_{0} x_{0}\Big) \right] - \frac{kT}{2} ~.
\end{eqnarray*}

Following an idea originally due to Feynman, we classify the paths in the functional integration in Eq.~(\ref{eq.p2i}) according to  their average point, defined as the functional
\begin{equation}
\bar x[x(t)] = \frac{1}{T} \int_{0}^T \-\-dt \,\,x(t)~,
\end{equation}
so that 
we can factor out in Eq.~(\ref{eq.p2i}) an ordinary integral over $\bar x$, namely
\begin{equation}\label{eq.PIADprice}
\rho(x_T, x_0, T) =  \int d\bar x  \,\, \rho_{\bar x}( x_T,  x_0, T ; \bar x)~, 
\end{equation}
where the {\em reduced density matrix}
\begin{align}\label{eq.reddens}
& \rho_{\bar x}(x_T,  x_0, T; \bar x)   =  \nonumber \\ &  \int_{x(0) =x_0}^{x(T) = x_T} \hspace{-0.1cm}{\cal D} [x(t)] \delta \left(\bar x - \frac{1}{T} \int_{0}^T dt \,\,x(t) \right) \, \,e^{S_{\rho}[x(t)]}~
\end{align}
represents the contribution to the path integral in  Eq.~(\ref{eq.p2i}) due to the paths that have $\bar x$ as their average point.

As the path integration has been reduced to paths belonging to the same class, we can develop a specialized approximation for each class. In particular, we can approximate the action in Eq.~(\ref{eq.action}) with a quadratic action in the 
displacement from the average point $\bar x$.  The original technique, independently developed in quantum physics  by  \cite{GiachettiTognetti1985} and \cite{FeynmanKleinert1986}, and dubbed GTFK, is limited to potentials with no explicit time dependence, {\em i.e.,} those originating from time independent drift and volatilities in the original SDE (\ref{eq.diffusion}). Here we generalize the GTFK approximation in order to employ a {\em time-dependent} quadratic potential of the form
\begin{equation}\label{eq.trialpot}
V_{\bar x}(x, t) = w(\bar x,t) + \gamma(\bar x, t) (x-\bar x) +  \frac{\omega^2(\bar x)}{2\sigma^2} (x-\bar x)^2~,
\end{equation}
where the functions $w(\bar x, t)$, $\gamma(\bar x, t)$  and $\omega^2(\bar x)$ are now to be optimized so that the {\em trial} reduced density matrix
\begin{align}\label{eq.trialreddens}
& \bar \rho_{\bar x}(x_T,  x_0, T; \bar{x})  \nonumber \\ &=  \int_{x(0) =x_0}^{x(T) = x} \hspace{-0.1cm}{\cal D} [x(t)] \delta \left(\bar x - \frac{1}{T} \int_{0}^T dt \,\,x(t) \right) \, \,e^{S_{\bar x}[x(t)]}~,
\end{align}
with the action given by
\begin{equation}\label{eq.trialaction}
S_{\bar x}[x(t)] = - \int_{0}^T dt \left[ \frac{1}{ 2\sigma^2 } \dot x^2(t) + V_{\bar x}(x(t), t) \right]~,
\end{equation}
best approximates the reduced density matrix in Eq.~(\ref{eq.reddens}).

The path integral in Eq.~(\ref{eq.trialreddens}), corresponding to the harmonic action (\ref{eq.trialaction}) is known in closed form in the general case that the parameters $k_{t}$, $\theta_{t}$, and $\sigma_{t}$ are time dependent \cite{stedman2024path}.  While the general solution is in terms of a second-order ODE, the solution is analytic in the case that both $k$ and
$\sigma$ are constant.  Specifically, we have
\begin{align}
&\bar{\rho}_{\bar x}\left(x_T, x_0, T; \bar{x}\right)= N(\bar x)   \times \nonumber \\ 
& \exp \left[-\frac{\left(x_{0}+x_{T}- 2(\bar{x} - \delta) \right)^2}{8 \alpha}-\frac{\omega \left(x_T-x_0\right)^2}{4\sigma^2} \coth f \right . \nonumber \\
& \left . -\left(x_T-\bar{x}\right) \frac{\Gamma_0}{\sinh 2 f}-\left(x_0-\bar{x}\right) \frac{\Gamma_T}{\sinh 2 f} \right]
\end{align}
with $f(\bar x) = \omega(\bar x) T/2$,
\begin{align}
N(\bar x) &= \sqrt{\frac{1}{2 \pi \alpha}\frac{1}{2 \pi T \sigma^2 }} \frac{f}{\sinh f}  e^{-\int_0^T dt \,w(\bar x, t) + \Gamma (\bar x)}~, \label{eq.N}  \\
\Gamma(\bar x) &=\frac{\sigma^2}{\omega}\left[\frac{\Gamma_{0T}}{\sinh 2 f}-\frac{1}{4 f}\left(\frac{\Gamma_0+\Gamma_T}{\sinh 2f} - \hat \gamma\right)^2\right]~, \\
\delta (\bar x) &= \frac{\sigma^2}{2\omega f} \left(\frac{\Gamma_0+\Gamma_T}{\sinh 2f} - \hat \gamma \right)~,  \\
\alpha(\bar x) &= \frac{\sigma^2}{2\omega}\left (\coth f -\frac{1}{f} \right )~, \label{eq.alpha}
\end{align}
and the following  definitions
\begin{align}\label{eq.gammas}
\Gamma_{0}(\bar x) &= \int_{0}^{T} dt \, \gamma(\bar x, t) \sinh \omega t~ \nonumber\\ 
\Gamma_{T}(\bar x) &= \int_{0}^{T} dt \, \gamma(\bar x, t) \sinh \omega (T- t)~  \nonumber \\
\Gamma_{0T}(\bar x) &= \int_{0}^{T} dt \, \gamma(\bar x, t) \sinh \omega (T- t) \nonumber \\ &\times \int_{0}^{t} ds \, \gamma(\bar x, s) \sinh \omega s~ \nonumber \\
\hat \gamma (\bar x) &= \int_{0}^{T}dt \,\gamma(\bar x,t)~.
\end{align}

In particular, the diagonal elements of the  reduced density matrix
\begin{align}\label{eq.reddens3}
& \bar \rho(x, x, T;  \bar x)_{\bar x}  \propto &\frac{1}{\sqrt{2\pi\alpha}} \exp\left[ -\frac{(x-\bar x +\delta_{\gamma})^2}{2\alpha} \right]~,
\end{align}
with
\begin{equation}\label{eq.deltagamma}
\delta_{\gamma}(\bar x) = \frac{\sigma^2}{2\omega} \left(  \frac{\Gamma_0+\Gamma_T}{2 \sinh^{2} f} - \frac{\hat \gamma}{f} \right)~, 
\end{equation}
take a suggestive form in terms of a Gaussian distribution with mean $\bar x$ and variance $\alpha(\bar x)$ that describes the fluctuations around an average point {\em shifted by} the quantity $\delta_{\gamma}$.  
We can use this distribution to determine the parameters $w(\bar x, t)$, $\gamma(\bar x, t)$,  and  $\omega(\bar x)$ in Eq.~(\ref{eq.trialpot}) by setting 
\begin{align} 
&\langle \langle V(x, t) \rangle\rangle =  \langle \langle V_{\bar x}(x,t) \rangle\rangle \nonumber \\ 
&= w(\bar x,t) -\gamma(\bar x, t)  \delta_{\gamma}(\bar x) + \frac{\omega^2(\bar x)(\alpha(\bar x)+ \delta_{\gamma}^{2}(\bar x))}{2\sigma^2}~,\nonumber \\ 
&\langle \langle V^{\prime}(x, t) \rangle\rangle =\langle \langle  V_{\bar x}^{\prime}(x,t)\rangle\rangle = \gamma(\bar x, t) - \frac{\omega^{2}(\bar x)}{\sigma^{2}}\delta_{\gamma}(\bar x)~,\nonumber\\
&\langle \langle V^{\prime\prime}(\bar x, t) \rangle\rangle =\langle \langle  V_{\bar x}^{\prime\prime}(\bar x, t)\rangle\rangle = \frac{\omega^2(\bar x)}{\sigma^2}~,\label{GTFK3}
\end{align}
with the short-hand notation 
\begin {align}
\langle \langle F(x) \rangle\rangle &\equiv \frac{1}{\sqrt{2 \pi\alpha(\bar x)}} \int_{-\infty}^{+\infty} d\xi \,\,e^{-\xi^2/2\alpha} F(\xi+\bar x - \delta_{\gamma}) \nonumber \\
 &= \left.e^{\frac{\alpha(\bar x)}{2}\partial_x^2} F(x)\right|_{x = \bar x-\delta_{\gamma}(\bar x)}~.
\end{align}
The equations above impose that the expectation value according to the Gaussian probability distribution in Eq.~(\ref{eq.reddens3}) of the drift potential, Eq.~(\ref{eq.driftpot}), and the quadratic approximation, Eq.~(\ref{eq.trialpot}), are in agreement with each other for {\em every} value of $\bar x$. 

Under the GTFK approximation, the value of a derivative Eq.~(\ref{eq.cc}) can be expressed as
\begin{align}\label{eq.cc2}
&V(0) = \nonumber \\ 
&\int  d\bar x  \, \int dx_T \,e^{-W(x_T, x_0)} \bar \rho_{\bar x}(x_T, x_0, T; \bar x) P(x_T)~.  
\end{align}
Remarkably,  the integral over $x_T$ can be performed analytically giving
\begin{equation}\label{eq.cc3}
V(0) = \int  d\bar x \, \bar N(\bar x)  e^{\frac{1}{4A(\bar x)} \partial^2_x} \left . P\left ( x \right)\right|_{x = x_{0}-\frac{B(\bar x)}{2A(\bar x)} }~,
\end{equation}
with 
\begin{align} 
A(\bar x) & =\frac{1}{8 \alpha}+\frac{\omega \coth f}{4 \sigma^2}+\frac{k}{2 \sigma^2}~, \nonumber \\ 
B(\bar x) & =\frac{\left(x_0-\bar{x}+\delta\right)}{2 \alpha}+\frac{\Gamma_0}{\sinh 2 f}+\frac{k\left(x_0-\theta_{T}\right)}{\sigma^2}~, \nonumber \\ 
C(\bar x) & =-\frac{\left(x_0-\bar{x}+\delta\right)^2}{2 \alpha}-\frac{\left(x_0-\bar{x}\right) (\Gamma_{0}+\Gamma_T)}{\sinh 2 f} \nonumber \\
&+\frac{k}{\sigma^2}(\theta_{T}-\theta_{0}) x_0 +\frac{k T}{2}~, \nonumber \\
\bar N(\bar x)  & = \sqrt{\frac{\pi}{A(\bar x)}} N(\bar x) \exp\left[C(\bar x) + \frac{B^2(\bar x)}{4A(\bar x)}\right]~,
\end{align}
so that option prices can be obtained with a {\em single} fast-converging integration when the application of the differential operator in Eq.~(\ref{eq.cc3}) gives rise to a finite set of terms.  This occurs for the calculation of survival probabilities, when $P\equiv 1$, and when the payoff can be approximated as a power series in $x$.


The GTFK approximation is, in fact, exact for harmonic actions. This is the case for generalized Gaussian models such as the Hull-White model \cite{hull1990pricing}, as shown in Ref.~\cite{stedman2024path}.



For the BK model, where $h(X_t) = \exp X_{t}$,  the GTFK conditions, in Eq. (\ref{GTFK3}), can be determined with some straightforward algebra as
\begin{align} 
\omega^2(\bar x) & =k^2+\sigma^2 \lambda \exp \left[\frac{\alpha}{2}+\bar{x}-\delta_\gamma\right] \nonumber \\
\gamma(\bar x, t)  & =\left(\frac{\omega^2-k^2}{\sigma^2}\right)\left(\delta_\gamma+1\right)+\frac{k^2 \bar{x}}{\sigma^2}+\frac{k \dot{\theta}_{t}}{\sigma^2}-\frac{k^2 \theta_{t}}{\sigma^2} \nonumber \\
w(\bar x, t) & =V\left(\bar{x}-\delta_\gamma\right)+\left(\frac{k^2-\omega^2}{2 \sigma^2}\right) \alpha-\frac{\omega^2 \delta_\gamma^2}{2 \sigma^2}\nonumber \\&+\delta_\gamma \gamma+\lambda\left(e^{\alpha / 2}-1\right) e^{\bar{x}-\delta_\gamma}~.
\label{eq:BKParams}
\end{align}

\begin{widetext}

\begin{table}[t]
{\begin{tabular}{@{}lccccc@{}}  \toprule
 $T$  &    ${\rm EE}$      & KL(1)                   &   KL(2)                 &   GTFK      &    PDE         \\ \colrule
0.1  & 611.87 (0.00) & 611.87 (0.00) & 611.87 (0.00) & 611.87 (0.00) & 611.87 \\
0.5  & 648.40 (0.00) & 648.40 (0.00) & 648.40 (0.00) & 648.40 (0.00) & 648.40 \\
1.0  & 692.43 (0.00) & 692.43 (0.00) & 692.43 (0.00) & 692.43 (0.00) & 692.43 \\
2.0  & 765.17 (0.58) & 765.76 (1.17) & 765.17 (0.58) & 764.59 (0.00) & 764.59 \\
3.0  & 809.03 (0.43) & 810.30 (1.70) & 809.45 (0.85) & 808.18 (0.42) & 808.60 \\
5.0  & 832.55 (0.91) & 836.49 (4.85) & 833.15 (1.51) & 830.42 (1.22) & 831.64 \\
10.0 & --            & 788.56 (17.02) & 776.31 (4.77) & 770.46 (1.08) & 771.54 \\
20.0 & --            & 705.29 (47.47) & 675.08 (17.26) & 659.88 (2.06) & 657.82 \\
\botrule
\end{tabular} }
\caption{{Black-Karasinski $T$ maturity credit spreads (in bps) obtained with the GTFK approximation, the Exponent Expansion (EE) of Ref.~\cite{stehlikova2014effective}, the Karhunen-Lo\'eve (KL) expansion of Ref.~\cite{daniluk2016} to first and second order, and by solving the associated PDE numerically. The parameters of the process, relevant for a stressed market scenario, are: mean-reversion speed  $k = 0.1$, level $\theta = \ln 0.04$, volatility $\sigma = 0.85$, and initial default rate $h_{0} = 0.06$. Absolute errors with respect to the PDE results are shown in parentheses.}}
\label{tablevsEE} 
\end{table}  

\end{widetext}

In the first equation in Eq.~(\ref{eq:BKParams}), $\omega(\bar x)$ appears in both the left-hand-side and, implicitly, through $\alpha$ and $\delta_{\gamma}$, as defined in Eqs.~(\ref{eq.alpha}) and (\ref{eq.deltagamma}), in the right hand side. The term $\delta_{\gamma}$ also couples the first two equations which, therefore, must be solved  simultaneously. In fact, it is more convenient to solve simultaneously for $\omega$ and $\delta_{\gamma}$, which are both {\em time-independent}, by using the first equation and the definition of $\delta_{\gamma}$, Eq.~(\ref{eq.deltagamma}), where the term $\hat \gamma$ is computed from the second equation in Eq.~(\ref{eq:BKParams}), namely
\begin{align}
\hat{\gamma}&=\left[\left(\frac{\omega^2-k^2}{\sigma^2}\right)\left(\delta_\gamma+1\right) +\frac{k^2 \bar{x}}{\sigma^2}\right] T \nonumber \\ &+ \frac{k}{\sigma^2}(\theta_{T}-\theta_{0})-\frac{k^2}{\sigma^2} \int_0^T dt \, \theta_{t}~.
\end{align}
The calculation generally converges in a few iterations, and is, therefore, very fast.  Once $\omega(\bar x)$ and $\delta_{\gamma}(\bar x)$ are found, $\gamma(\bar x, t)$ and $w(\bar x, t)$ can be computed directly.

As an illustration of the accuracy of the GTFK approach we have compared the results for survival probabilities $Q(x_{0}, T)$, in Eq.~(\ref{eq.survprob}) for a wide range of maturities, $T$, and different volatilities and time-dependent mean reversion functions, $\theta_{t}$. Here, as it usually the case in practice, we consider piecewise constant $\theta_{t}$ functions. 

In Fig.~\ref{zeroBK}, we compare the results of the GTFK approximation with those obtained by solving the appropriate PDE with a finite difference scheme.  The results presented, representative of the general case, are for the particular choice
$
\theta_t= \theta_{min} + {(\theta_{max}-\theta_{min})}/{N_{l}} ~\eta(t)~,
$
where $N_{l}$ is the number of levels we use to discretize the range $[\theta_{min},\theta_{max}]$  and $\eta(t) = \sum_{i=0}^{N_{l}-1} {\bm 1}(t \ge t_{i})$, with $t_{i} = i T / N_{l}$.
We see that the GTFK approximation provides extremely accurate results up to very large maturities and volatilities. 

\begin{table}[b]
{\begin{tabular}{@{}lcccc@{}}  \toprule
${T}$ & HJT(1) & HJT(2) & GTFK & PDE \\
\hline
1.1 & 0.9940 (0.02\%) & 0.9938 (0.04\%) & 0.9942 (0.00\%) & 0.9942 \\
1.5 & 0.9699 (0.07\%) & 0.9684 (0.09\%) & 0.9693 (0.00\%) & 0.9693 \\
2.0 & 0.9398 (0.5\%)  & 0.9356 (0.02\%) & 0.9354 (0.001\%) & 0.9354 \\
3.0 & 0.8800 (2.0\%)  & 0.8661 (0.4\%)  & 0.8625 (0.01\%)  & 0.8624 \\
4.0 & 0.8214 (4.0\%)  & 0.7916 (0.2\%)  & 0.7904 (0.03\%)  & 0.7901 \\
5.0 & 0.7648 (5.6\%)  & 0.7133 (1.5\%)  & 0.7247 (0.05\%)  & 0.7243 \\
\botrule
\end{tabular}}
\caption{Black-Karasinski forward survival probabilities obtained with the GTFK approximation, the approach of Ref.~\cite{horvath2018analytic} (HJT)  to first and second order, and by solving the associated PDE numerically, for the forward survival probabilities $Q(x_{t}, t, T)$ for $t=1$ and different values of $T$. The models were calibrated to a $5$Y spread in the same stressed market environment considered in Table \ref{tablevsEE} but with a time-dependent mean reversion level
with $\theta_{min} = \log(0.04)$ and $\theta_{max} = \log(0.048)$ and $N_{l}=2$. Relative errors (in absolute value) with respect to the PDE solution are shown in parentheses. }
\label{horvath2018}
\end{table}

{Table \ref{tablevsEE} illustrates that, in the time-homogeneous case, the GTFK method compares favorably with the results obtained with other semi-analytical approximations, namely the Exponent Expansion (EE) \cite{stehlikova2014effective}, and the  Karhunen-Lo\'eve (KL) expansions \cite{daniluk2016} when benchmarked against a numerical solution of the associated PDE. We have considered parameters that are relevant for periods of stress, {\em e.g.}, with an implied 3-months outturn volatility $\sim 84\%$. For short time horizons, all methods have comparable accuracy. For longer time horizons, the GTFK remains very accurate while the KL deteriorates significantly and the EE, which has a finite convergence ratio in $T$, eventually breaks down.}

{In Table \ref{horvath2018}, we compare the GTFK results against those of Ref.~\cite{horvath2018analytic} for  {\em forward-starting} survival probabilities 
\begin{equation}
Q(x_{t}, t, T) = \mathbb{E}\left[1_{\{\tau>T\}}| x_{t}, \tau > t\right]=\mathbb{E}_{t}\left[e^{-\int_t^T du\, h_{u}}\right]~,
\end{equation}
for $t=1$ and different values of $T$, for a model calibrated to a $5$Y spread in the same stressed market environment considered in the previous table but with a time-dependent mean reversion level
with $\theta_{min} = \log(0.04)$, $\theta_{max} = \log(0.048)$, and $N_{l}=2$. Here, $x_{t}$ is the one implied by the instantaneous forward curve. This confirms the accuracy of the GTFK approximation.}

\begin{widetext}

\begin{table}[t]
\begin{tabular}{ccccc|cccc}
  \toprule
       & \multicolumn{4}{c|}{Normal Markets} & \multicolumn{4}{c}{Stressed Markets} \\
  $T$  & GTFK   & PDE    & Cal. Target & Abs.~Diff. & GTFK   & PDE    & Cal. Target  & Abs.~Diff. \\
  \colrule
  1.0   & 39.67   & 39.67  & -          & 0.00    & 482.32 & 482.32 & -          & 0.00 \\
  2.0   & 40.13  & 40.13   & -         & 0.00     & 521.39 & 521.37 & -          & 0.02 \\
  3.0   & 40.33   & 40.33  & 40.33  & 0.00     & 540.96 & 541.00 & 541.00 & 0.04 \\
  4.0   & 50.19   & 50.16  & -         & 0.03     & 549.7   & 549.8 & -          &  0.1 \\
  5.0   & 71.3    & 70.3   & 70.3     & 1.0       & 552.9   & 553.00 & 553.00 & 0.1 \\
  7.0   & 99.6    & 100.1 & -          & 0.5       &  548.6  & 548.5  & -          & 0.1 \\
  10.0 & 119.8   & 122.1 & 122.1   & 2.3       &  537.3  & 537.0   & 537.0  & 0.3 \\
  20.0 & 143.2   & 147.5 &  -         & 4.3       & 519.6   & 518.9  & -          & 0.7 \\
\botrule
\end{tabular}
\caption{{GTFK $T$-maturity credit spreads (in bps) obtained by calibrating the discrete levels of the mean reversion rate for the Black-Karasinski model to the 3, 5, 10 years CDS spreads via PDE for the Italian sovereign CDS in normal (as represented by the current 52-week average) and stressed (as of January 4th, 2012) market conditions. The other parameters are $k=1$ ({\em resp.}, 0.5), and $\sigma =0.5$ ({\em resp.}, 0.95), for the normal ({\em resp.}, stressed) market conditions.}}
\label{CDSspreads}
\end{table}

\end{widetext}

{In Table \ref{CDSspreads} we compare the GTFK results with those obtained with the numerical solution of the PDE for the BK model calibrated to mid Credit Default Swap (CDS) spreads on Republic of Italy at the 3Y, 5Y and 10Y tenors. We consider two market regimes: a normal one represented by the current 52-week average and a stressed one, corresponding to that observed on January 4th, 2012, during the European sovereign debt crisis. We observe that, even when the mean reversion level is strongly time dependent, {\em e.g.}, as required to calibrate to a steeply increasing term structure of credit spreads in normal markets, the GTFK approximation produces accurate results well within typical bid-ask spreads. In stressed markets, when the term structure of credit spreads is typically flatter, the GTFK approximation is even more accurate despite the much higher volatility.} 




From a computational perspective, within the GTFK method the calculation of the AD prices, Eq. (\ref{eq.ADprice}), or survival probabilities, see {\em e.g.}, Eq. (\ref{eq.cc3}) with $P\equiv 1$, involves a single integration over the average point $\bar x$ of a well-behaved integrand, decaying exponentially fast.
Indeed, for the case that $\theta_{t}$ is a piecewise constant function with $N_{l}$
intervals, as is typical of the output of a model calibration routine, all of the
integrals in Eqs.~(\ref{eq.N}) and (\ref{eq.gammas}) can be performed analytically for each
interval.  As a result, each one-dimensional integral requires $O(N_{l})$ operations, while
the two-dimensional integral, $\Gamma_{0T}(\bar x)$, requires $O(N_{l}^{2})$ operations.
Therefore, the overall complexity of the GTFK calculation of a survival probability is
$O(N_{\bar x}N_{l}^{2})$, where $N_{\bar x}$ is the number of function evaluations in the quadrature over $\bar x$. {For comparison, finite-difference methods for PDEs have a computational complexity of $O(N_{x}N_{t})$, where $N_{x}$ and $N_{t}$ are the number of points in the space and time direction, respectively.} 

In practice, we set $N_{l}$ to the number of traded CDS maturities for a single
issuer, typically $< 10$, and find good convergence using a standard 41-point quadrature
scheme.  Hence, the GTFK approximation compares favorably with both the computational
complexity and memory requirements of finite-difference methods for PDEs, 
especially for longer time horizons. {For example, for maturity $T = 5$, our current implementation of the GTFK approximation is about 10 times faster than our PDE solver with grid size $N_{x} = N_{t} = 200$, and could be optimized further.} Furthermore, it is not affected by the common numerical instabilities that such discretization schemes carry, especially for the computation of AD densities. We also note that the low memory requirements of the GTFK approximation make it particularly attractive for the calculation of XVA on a large portfolio of CDS since
it enables the calculation of conditional default probabilities within the XVA simulation without the storage overhead of a PDE grid for each reference entity.

As a final example, we consider the pricing of a quanto CDS,
in which the running premium and protection payment upon default are made in a non-standard or ``foreign'' currency. 
{It is possible to relate the credit spreads in the domestic and foreign currency by introducing a modeling framework which captures the correlation between the default intensity and the FX rate.  In particular, following \citep{brigo2019multi}, it is important to introduce a devaluation mechanism of the FX rate upon default, {\em e.g.}, with a process for the FX rate of the form
\begin{equation}
  dY_t=\mu_{Y} Y_t dt + \sigma_Y Y_t dW_t^Y + J Y_{t-} d N_t~,
\end{equation}
where $Y_t$ is the value at time $t$ of one unit of foreign currency in the domestic currency, $N_{t}$ is a default process with intensity $\int_{0}^{t} du \,h_{u} $, and \(J\) is the constant proportional jump size. We also assume that the Brownian motions \(W_{t}^Y\) and \(W_{t}\) are correlated with a constant correlation coefficient \(\rho\).
}

\begin{table}[t]
\begin{tabular}{cccc}
  \toprule
  $T$  & GTFK   & PDE    & Abs.~Diff.  \\
  \colrule
  1.0   & 79.80  & 79.81 & 0.01 \\
  2.0   & 88.67  & 88.68   & 0.01 \\
  3.0   & 93.37  & 93.41  & 0.04 \\ 
  4.0.  & 95.77  & 95.85  & 0.08 \\
  5.0   & 96.9    & 97.0  & 0.1 \\
  7.0   & 96.9    & 97.0  & 0.1 \\
  10.0 & 95.5    & 95.8  & 0.3\\
  20.0 & 93.8    & 93.6  & 0.2 \\  
\botrule
\end{tabular}
\caption{{GTFK $T$-maturity EUR-USD quanto CDS basis  (in bps) obtained with the stressed parameters used in  Table \ref{CDSspreads}, with $\rho = -0.2$, $\sigma_{Y}=0.13$ and calibrating $J$ to the quanto CDS basis at the 5Y maturity, obtained by PDE, during the European sovereign debt crisis (97 bps).}}
\label{table.quantoCDS}
\end{table}

Using measure-change results for jump diffusions, see {\em e.g.}, \cite{brigo2019multi}, we can show that the intensity $h^f_{t}$ of $N_{t}$ under the foreign currency risk-neutral measure $Q_f$ is given by $h_{t}^{f} = (1+J) \exp X_{t}^{f}$ with 
\begin{equation}
dX_t^f=\kappa \left(\theta^{f}_{t}-X_{t}^{f}\right) dt+\sigma dW_t^{f}~,
\end{equation}
$X_{0}^{f} = X_0$, where $W_{t}^{f}$ is a standard Brownian motion under $Q_f$ and
\begin{equation} \label{eq.quantomeanrev}
\theta^{f}_{t}= \theta_{t} + \rho \sigma \sigma_Y~.
\end{equation}
Therefore, under the foreign risk neutral measure $Q_f$, the intensity of default is still given by a BK process, with the same volatility but with a different mean reversion level and an overall scaling factor. 

{In the same comprehensive study, \cite{brigo2019multi} showed that the model considered here is rich enough to calibrate to market quotes for the domestic and foreign currency CDS spreads even in regimes of stress, such as the European sovereign debt crisis of 2011 and 2012. In that study, the authors relied on a fully numerical solution of the pricing PDE for the calculation of CDS spreads. Here we also calculate survival probabilities using our GTFK approximation for the BK dynamics. In particular, for the foreign currency measure, we can set $\lambda = 1+J$ and employ the simple modification of the  mean reversion function in Eq.~(\ref{eq.quantomeanrev}). The results obtained, for a set of parameters relevant for the 2011-2012 sovereign debt crisis, are shown in Table.~\ref{table.quantoCDS}.  They demonstrate that the GTFK approximation provides results which are well within the bid-ask spread of those obtained by solving the PDE numerically. This confirms the practical utility of the GTFK approximation as an alternative to fully numerical implementations.} 


In conclusion,  we have presented, the GTFK method, a path-integral approach for the calculation of transition  probabilities, Arrow-Debreu densities,  and option prices for a general class of default intensity models.  


The accuracy and ease of computation of the GTFK method makes it a computationally efficient alternative to PDE or Monte Carlo. 
This is of practical utility in a variety of derivatives pricing applications,
such as the calculation of conditional discount factors and survival probabilities in the context of multifactor simulations, {\em e.g.}, for XVA of credit products.
Furthermore, although we have presented the GTFK approximation 
in the context of default intensity models, the method could be applied to other problems such as
Asian options, options on realized variance or goal-based wealth management optimization  \cite{halperin2024scop}.

It is a pleasure to acknowledge stimulating discussions with Jacky Lee, Colin Turfus and Sebastian Jaimungal. We thank Tianci Xu, and  Zhe Fei for helping with the implementation of the HJT approach. The authors are grateful to the referees for their valuable suggestions.

\appendix

\bibliography{biblio}     

\end{document}